\documentclass[aps,twocolumn,showpacs,floats,prl]{revtex4}
\usepackage{graphicx}
\usepackage{dcolumn}
\usepackage{bm}
\usepackage{color}
\usepackage{subfigure}
\usepackage{amsmath}

\begin{document}

\author{L. Parisi$^1$, G. E. Astrakharchik$^2$, and S. Giorgini$^1$}

\affiliation{
$^1$ Dipartimento di Fisica, Universit\`a di Trento and CNR-INO BEC Center, I-38050 Povo, Trento, Italy \\
$^2$ Departament de F\'{i}sica i Enginyeria Nuclear, Universitat Polit\`ecnica de Catalunya, Campus Nord B4-B5, E-08034, Barcelona, Spain
}

\title{The liquid state of one-dimensional Bose mixtures: a quantum Monte-Carlo study} 

\begin{abstract} 
By using exact quantum Monte-Carlo methods we calculate the ground-state properties of the liquid phase in one-dimensional Bose mixtures with contact interactions. We find that the liquid state can be formed if the ratio of coupling strengths between inter-species attractive and intra-species repulsive interactions exceeds a critical value. As a function of this ratio we determine the density where the energy per particle has a minimum and the one where the compressibility diverges, thereby identifying the equilibrium density and the spinodal point in the phase diagram of the homogeneous liquid. Furthermore, in the stable liquid state, we calculate the chemical potential, the speed of sound, as well as structural and coherence properties such as the pair correlation function, the static structure factor and the one-body density matrix, thus providing a detailed description of the bulk region in self-bound droplets. 
\end{abstract}
\maketitle 

Ultracold atoms provide a rich toolbox to realize different states of matter where many-body correlations can be investigated in a very clean experimental setup. In early years, the most common gas phase, both in the normal and superfluid regime, and artificial crystals created by external lattice potentials were routinely produced~\cite{RMPBloch}. More recently, interaction effects have been exploited to obtain spontaneous breaking of translational symmetry in ordered arrangements of particles subject to long-range forces~\cite{Rydberg} and in most exotic supersolid systems, featuring both the rigidity of standard solids and the dissipationless motion of vacancies typical of superfluids~\cite{Supersolid1, Supersolid2}. Furthermore, self-bound liquid droplets were generated as a result of quantum fluctuations in samples interacting via anisotropic dipolar forces~\cite{Pfau1, Pfau2, Pfau3, Pfau4, Ferlaino} as well as via contact interparticle potentials~\cite{Tarruell1, Tarruell2, Fattori17}. 

In three and two dimensions such droplets would collapse according to mean-field theory and are stabilized, for large enough numbers of particles, by repulsive correlations beyond the mean-field description. Dipolar droplets were characterized theoretically by means of a generalized nonlocal, nonlinear Schr\"odinger equation~\cite{Santos16}  and also by exact quantum Monte-Carlo (QMC) methods, employing a model two-body potential with hard-core repulsion~\cite{Boronat17}. Droplets in a two-component Bose gas with short-range interactions have been first predicted and studied using a generalized Gross-Pitaevskii (GGP) equation in Ref.~\cite{Petrov15}. QMC simulations of these latter systems have also been carried out, even though only for limited numbers of particles~\cite{Boronat18}. 
 
In one spatial dimension (1D), quantum droplets of Bose mixtures with contact interactions have been predicted to occur as a result of a different mechanism. Here, beyond mean-field fluctuations are attractive and one needs a net mean-field repulsion in order to stabilize the droplet, which therefore are expected to form in the region where, according to mean-field theory, the homogeneous gas mixture is still stable~\cite{Petrov16}. The approach based on the GGP equation is valid in the weak-coupling limit and provides a full description of the ground-state energetics of the bulk liquid phase as well as of the density profiles in droplets with a finite number of particles~\cite{Petrov16}. 

Droplets in 1D are also particularly interesting because of the enhanced role of quantum fluctuations and because stable regimes of strong correlations are experimentally achievable~\cite{Stoferle04, Paredes04, Weiss05, Haller09} and enjoy enhanced stability. This opens the intriguing perspective of investigating the 1D liquid phase when interactions are strong and can not be accounted for by the GGP approach. In the present Letter we address theoretically the regime of strongly correlated liquids by means of exact QMC methods applied to a 1D mixture of Bose gases with contact interactions. We determine the phase diagram of the homogeneous liquid in terms of density and coupling strengths. Furthermore, in bulk systems at equilibrium a number of relevant thermodynamic quantities is calculated, such as chemical potential and compressibility, as well as the behaviour of correlation functions which provides a clear indication of the presence of strong interactions.

We consider the following Hamiltonian
\begin{eqnarray}
H&=&-\frac{\hbar^2}{2m}\sum_{i=1}^{N_a} \frac{\partial^2}{\partial x_i^2}+g\sum_{i<j}\delta(x_i-x_j) - \frac{\hbar^2}{2m}\sum_{\alpha=1}^{N_b}\frac{\partial^2}{\partial x_\alpha^2}
\nonumber
\\
&+& g\sum_{\alpha<\beta}\delta(x_\alpha-x_\beta)+\tilde{g}\sum_{i,\alpha}\delta(x_i-x_\alpha) \;,
\label{Hamiltonian}
\end{eqnarray}
composed of the kinetic energy of the two components with the same mass $m$ and atom numbers $N_a$ and $N_b$, of the repulsive intra-species potentials modelled by the same coupling constant $g>0$ and by the attractive inter-species potential of strength $\tilde{g}<0$. Here $x_i$ with $i=1,\dots,N_a$ and $x_\alpha$ with $\alpha=1,\dots,N_b$ denote, respectively, the positions of particles belonging to component $a$ and $b$ of the mixture. In a box of size $L$ the homogeneous densities of the two components are given by $n_a=N_a/L$ and $n_b=N_b/L$. We consider balanced systems where $N_a=N_b=N/2$, such that the relevant dimensionless coupling parameters are given by $\gamma=\frac{gm}{n\hbar^2}$ and $\eta=\frac{|\tilde{g}|m}{n\hbar^2}$ in terms of the total density $n=N/L$. An important energy scale is fixed by the binding energy of dimers in vacuum, $\epsilon_b=-\frac{\hbar^2}{m\tilde{a}^2}$, where $\tilde{a}=\frac{2\hbar^2}{m|\tilde{g}|}$ is the 1D scattering length associated with the attractive inter-species contact potential.

Let us first discuss the ground state of the Hamiltonian in Eq.~(\ref{Hamiltonian}) in the weak-coupling limit, corresponding to $\gamma\ll1$ and $\eta\ll1$. The energy density in terms of the total density $n$ is given by
\begin{eqnarray}
\frac{E_{\text{GGP}}}{L}&=&\frac{n^2}{4}\left(g-|\tilde{g}|\right)
\nonumber
\\
&-&\frac{\sqrt{m}n^{3/2}}{3\sqrt{2}\pi\hbar}\left[\left(g-|\tilde{g}|\right)^{3/2}+\left(g+|\tilde{g}|\right)^{3/2}\right] \;.
\label{EnerGGP}
\end{eqnarray}
This represents the local energy to which the GGP functional adds the kinetic energy contribution $\frac{\hbar^2}{2m}(\nabla\sqrt{n})^2$~\cite{Petrov16}. From the ground-state energy $E$ of the mixture one can extract all relevant thermodynamic quantities: the extremum condition $\frac{d E/N}{d n}=0$ yields the equilibrium density $n_{\text{eq}}$ of the liquid and the relations $\mu=\frac{dE}{dN}$ and $mc^2=n\frac{d\mu}{dn}$ calculated at the density $n_{\text{eq}}$ give, respectively, the chemical potential $\mu_{\text{eq}}$ and the speed of sound $c_{\text{eq}}$ at equilibrium. In the GGP approach these quantities are obtained using $E_{\text{GGP}}$ of Eq.~(\ref{EnerGGP}) as a perturbative approximation to the energy $E$.

We study the ground-state properties of the Hamiltonian in Eq.~(\ref{Hamiltonian}) in a box of size $L$ with periodic boundary conditions by means of QMC techniques. More specifically, the diffusion Monte-Carlo (DMC) method solves the many-body Schr\"odinger equation in imaginary time, thereby obtaining the exact ground-state energy through a large-time projection~\cite{DMC}. Importance sampling is implemented via a guiding function, which also encodes the contact boundary conditions imposed by the interactions in the Hamiltonian. The guiding wave function is constructed as a product of pairwise correlation terms which, at short interparticle distance, reproduce the exact solution of the two-body problem with the contact potential and at longer distances account for many-body correlations~\cite{Parisi, SuppMat}. Finite-size effects are considered by performing calculations with different $N$ and are found to be smaller than the typical statistical uncertainty.

\begin{figure}
\begin{center}
\includegraphics[width=8.0cm]{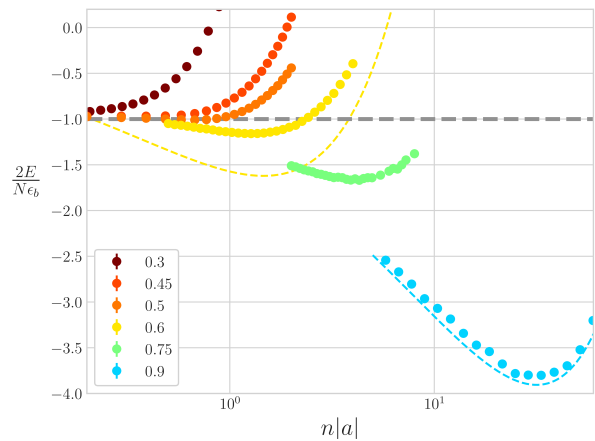}
\caption{Energy per particle, in units of half of the binding energy, as a function of the density for different values of the ratio $|\tilde{g}|/g$ of coupling constants. Error bars are smaller than the symbol sizes. The dashed line is the result of the GGP approach at $|\tilde{g}|/g=0.9$ (cyan) and at $|\tilde{g}|/g=0.6$ (yellow).}
\label{fig1}
\end{center}
\end{figure}

\begin{figure}
 \begin{center}
\includegraphics[width=8.0cm]{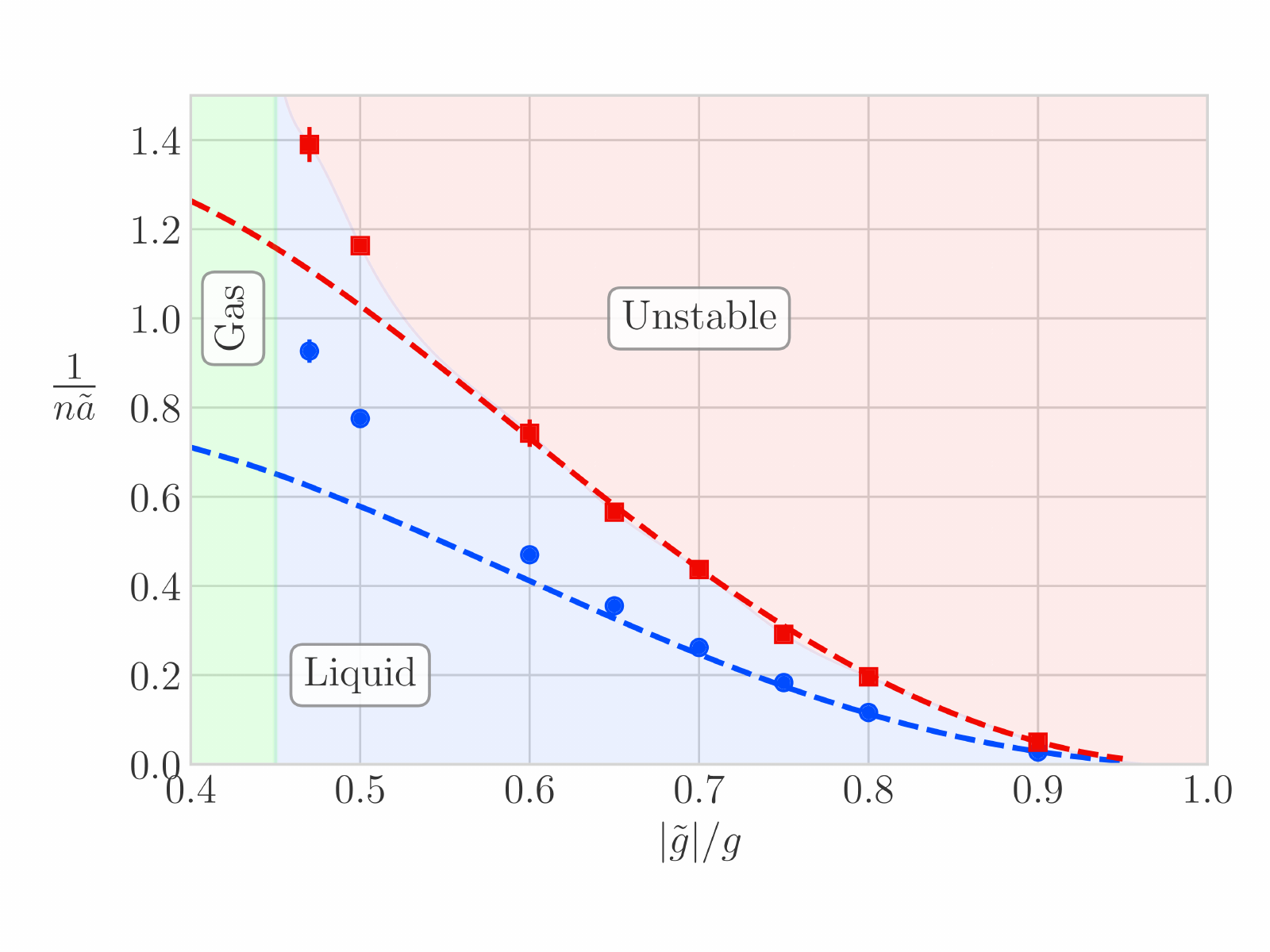}
\caption{Phase diagram of the homogeneous liquid phase: (blue) circles correspond to the equilibrium density of the liquid and (red) squares to the spinodal point where the compressibility diverges. Dashed lines refer to the predictions of the GGP theory.}
\label{fig2}
\end{center}
\end{figure}

The results for the ground-state energy per particle $E/N$ are shown in Fig.~\ref{fig1} for fixed values of the ratio of coupling constants $|\tilde{g}|/g$ and as a function of the dimensionless gas parameter $n|a|$. Here, $a=-\frac{2\hbar^2}{mg}$ is the scattering length associated with collision processes of the repulsive intra-species potential with strength $g$. Two distinct behaviours are clearly visible: if the ratio $|\tilde{g}|/g$ is sufficiently small the energy is a monotonously increasing function of the density signalling a gas phase where the minimum of energy is reached at a vanishing density and corresponds to half of the binding energy $\epsilon_b$. On the contrary, if the ratio is larger than a critical value, a minimum shows up in $E/N$ and the density at the minimum corresponds to the equilibrium density of the liquid phase. The critical ratio of coupling strengths is found to be $(|\tilde{g}|/g)_{\text{crit}}=0.47(2)$. This value is in close agreement with the result of the four-body scattering problem where the effective interaction between dimers crosses from repulsive to attractive~\cite{Petrov18}. Simultaneous effective three-dimer repulsion~\cite{Guijarro18} provides a microscopic scenario for the formation of the liquid which is consistent with our many-body calculations. Fig.~\ref{fig1} reports also the result of the GGP theory based on the energy functional of Eq.~(\ref{EnerGGP}). At high density, where the weak-coupling theory is applicable, we find good agreement, but large deviations both in the energy of the minimum and in the shape of curve are visible at small density. Similar results for the 3D homogeneous liquid phase have been obtained in Ref.~\cite{Zillich18} using a variational approach.

The curves shown in Fig.~\ref{fig1} allow us to determine the phase diagram of the homogeneous liquid in the region of ratios $(|\tilde{g}|/g)_{\text{crit}}<|\tilde{g}|/g<1$ where this state  can exist. The phase diagram is shown in Fig.~\ref{fig2}, where we report the values of the equilibrium density $n_{\text{eq}}$ and of the spinodal density, defined as the point where $\frac{d^2 E/L}{dn^2}=0$. At density $n$ below the spinodal line the homogeneous system is mechanically unstable and breaks into droplets. For larger values of $n$ the homogeneous phase is stable with a positive or negative pressure depending on whether $n$ is larger or smaller than $n_{\text{eq}}$. We also find that the GGP approach is quite reliable in predicting both the equilibrium and the spinodal line. Deviations start to appear for $|\tilde{g}|/g\lesssim0.6$.    

\begin{figure}
\begin{center}
\includegraphics[width=8.0cm]{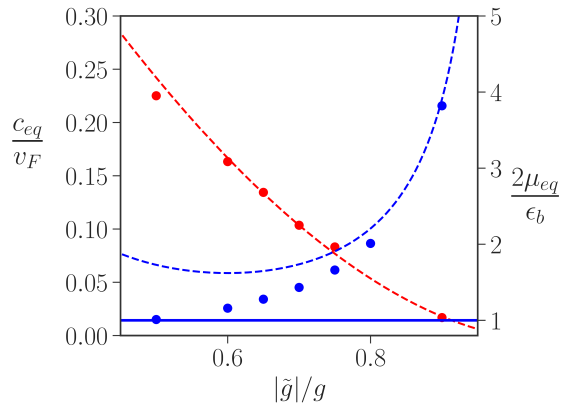}
\caption{Chemical potential $\mu_{\text{eq}}$ in units of half of the dimer binding energy: (blue) circles and right vertical axis. Speed of sound $c_{\text{eq}}$ in units of the Fermi velocity: (red) squares and left vertical axis. The dashed lines correspond to the results of the GGP theory.}
\label{fig3}
\end{center}
\end{figure}

Various ground-state properties of the liquid state at the equilibrium density $n_{\text{eq}}$ are shown in Fig.~\ref{fig3} as a function of the ratio $|\tilde{g}|/g$. In particular, we provide results for the chemical potential $\mu_{\text{eq}}$, which determines the rate of evaporation of particles from a droplet due to thermal effects, and the speed of sound $c_{\text{eq}}$, fixing the low-lying collective modes of the droplet. Significant deviations compared to the GGP approach are found for $\mu_{\text{eq}}$ at small ratios, where the weak-coupling theory fails to recover the physics of bound dimers. On the other hand, we find that $c_{\text{eq}}$ is well described by the GGP energy functional down to the smallest values of $|\tilde{g}|/g$ considered in Fig.~\ref{fig3}. Notice, however, that the speed of sound is reported here in units of the Fermi velocity $v_F=\frac{\hbar\pi n_{\text{eq}}}{2m}$ which itself depends on the equilibrium density $n_{\text{eq}}$.

\begin{figure}
\begin{center}
\includegraphics[width=8.0cm]{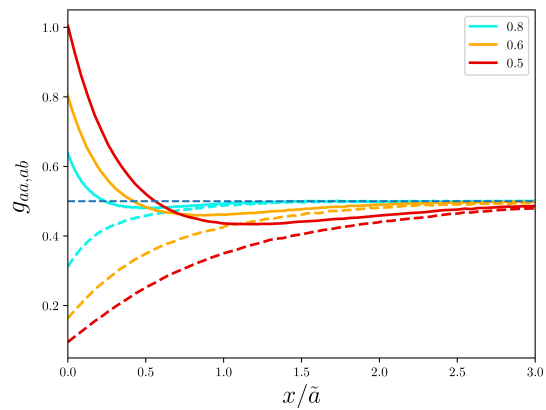}
\caption{Pair correlation function of parallel spins $g_{aa}$ (dashed lines) and anti-parallel spins $g_{ab}$ (solid lines) in the liquid for different values of the ratio $|\tilde{g}|/g$ of coupling constants.}
\label{fig4}
\end{center}
\end{figure}

\begin{figure}
\begin{center}
\includegraphics[width=8.0cm]{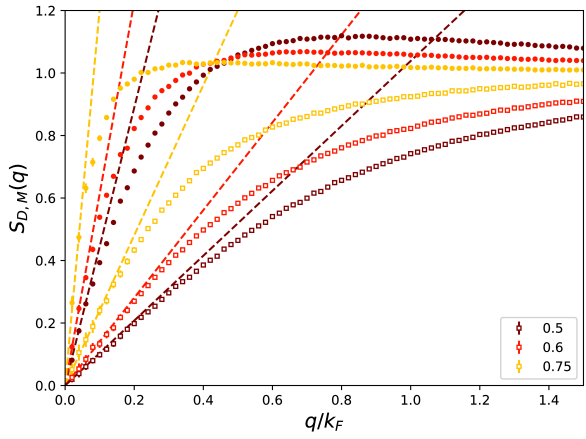}
\caption{Density (full symbols) and magnetic (open symbols) static structure factor in the liquid as a function of $q/k_F$ for different values of the ratio $|\tilde{g}|/g$. Here $k_F=\frac{\hbar\pi n}{2}$ is the Fermi wave vector. Dashed lines correspond to the low-$q$ linear dependence fixed by the compressibility and by the magnetic susceptibility respectively for $S_D(q)$ and $S_M(q)$.}
\label{fig5}
\end{center}
\end{figure}

\begin{figure}
\begin{center}
\includegraphics[width=8.0cm]{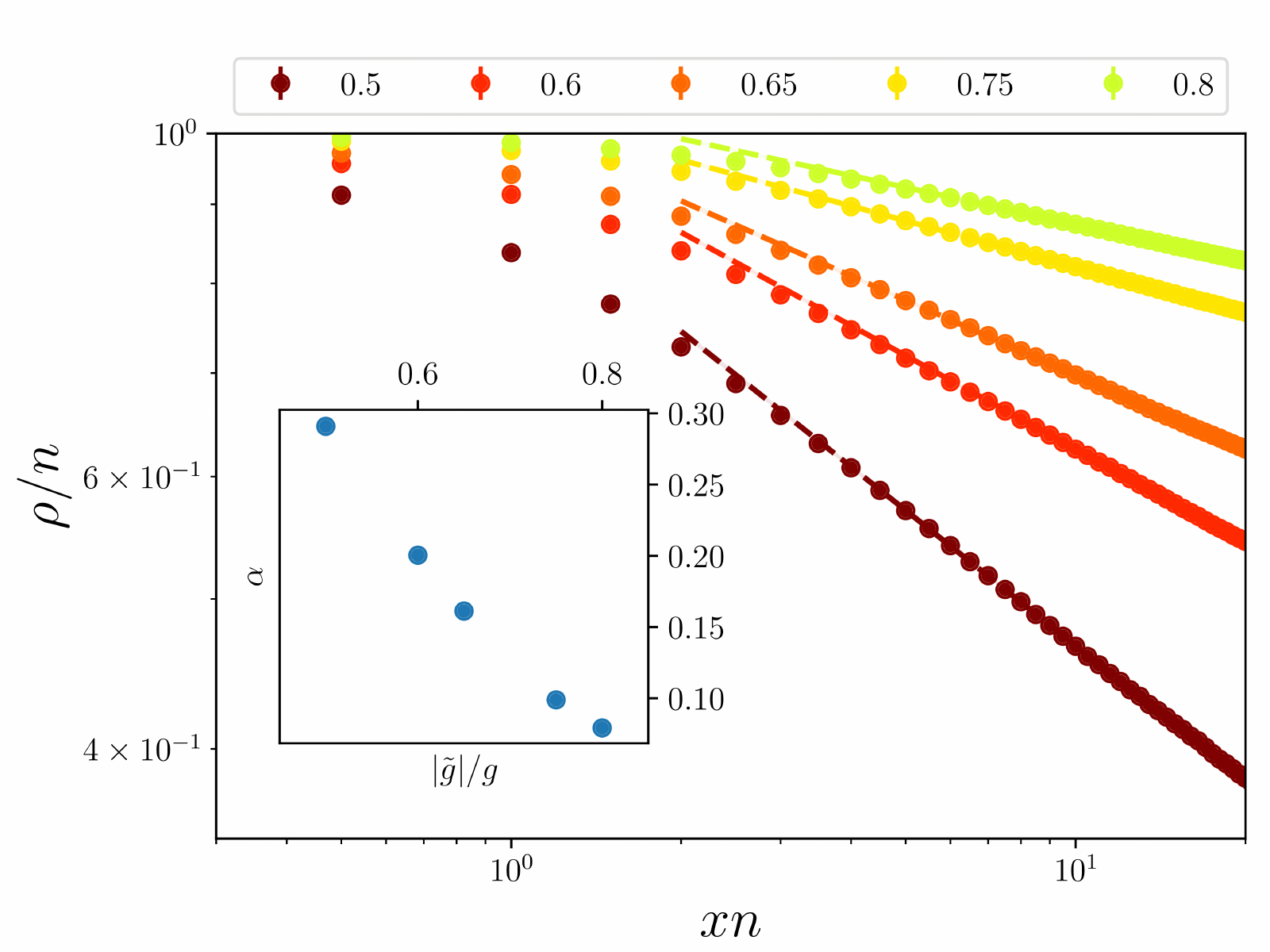}
\caption{Spatial dependence of the OBDM for different values of the ratio $|\tilde{g}|/g$ of coupling constants. Dashed lines are power-law fits $1/x^\alpha$ to the long-range behaviour. In the inset we report the values of the exponent $\alpha$ obtained from the fit.}
\label{fig6}
\end{center}
\end{figure}

Relevant information about the structure of the liquid state at equilibrium are obtained from the study of correlation functions. The pair correlation functions of parallel and anti-parallel spins are defined as expectation values $\langle\dots\rangle$ over the ground state
\begin{eqnarray}
g_{aa}(s)&=&1+\frac{4}{n^2}\left( \langle\delta n_a(x+s)\delta n_a(x)\rangle-\frac{n}{2}\delta(s)\right)
\nonumber\\
g_{ab}(s)&=&1+\frac{4}{n^2}\langle\delta n_a(x+s)\delta n_b(x)\rangle \;,
\label{pairfun}
\end{eqnarray} 
of the density fluctuations $\delta n_a(x)=\sum_{i=1}^{N_a}\delta(x-x_i)-\frac{n}{2}$ and $\delta n_b(x)=\sum_{\alpha=1}^{N_b}\delta(x-x_\alpha)-\frac{n}{2}$ of the two components measured with respect to the average density. These functions are shown in Fig.~\ref{fig4} for different values of the ratio $|\tilde{g}|/g$. At large distances correlations vanish yielding the result $g_{aa}=g_{ab}=1$. The anti-parallel spin correlation function $g_{ab}$ shows a long-range suppression and a peak for $s\lesssim\tilde{a}$. This behaviour arises from the short-range pairing between opposite spins occurring on length scales of the order of the size $\tilde{a}$ of a dimer and from the phononic long-range tail. By reducing the ratio $|\tilde{g}|/g$ both the minimum and the height of the peak become more prominent. On the contrary, the behaviour of $g_{aa}$ is fully determined by the repulsive intra-species correlations and it exhibits a monotonously decreasing behaviour as the distance is reduced. Also in this case, for smaller values of $|\tilde{g}|/g$, correlation effects are stronger and close to the critical ratio the repulsion between like particles produces a large suppression of $g_{aa}$. We notice that the density pair correlation function, defined as the average $g_{\small D}(s)=\frac{1}{2}[g_{aa}(s)+g_{ab}(s)]$, is peaked at short distances signalling the dominant role of attractive interactions characteristic of a liquid. Conversely, the magnetic pair correlation function defined as $g_{\small M}(s)=1+\frac{1}{2}[g_{aa}(s)-g_{ab}(s)]$ is suppressed at short distances as a consequence of the repulsion between dimers. 

From the Fourier transforms of the pair correlation functions $g_{\small{D(M)}}(s)$ one obtains the density and magnetic static structure factors defined as $S_{\small{D(M)}}(q)=1+n\int ds\; e^{iqs}\left(g_{\small{D(M)}}(s)-1\right)$. Both structure factors are shown in Fig.~\ref{fig5}. At large momenta $S_{\small D}(q)$ and $S_{\small M}(q)$ tend to unity, while for small values of $q$ we find in both cases a linear dependence. This is expected in the case of the density structure factor which should obey the law $S_{\small D}(q)=\frac{\hbar q}{2mc_{\text{eq}}}$ fixed by the speed of sound $c_{\text{eq}}$. In the case of $S_M(q)$, instead, one might expect a quadratic dependence as $q\to0$ caused by the presence of a pairing gap in the spin sector~\cite{Yang-Gaudin}. However, as evident from Fig.~\ref{fig3}, we are in the regime $|\epsilon_b|\lesssim|\mu_{\text{eq}}|$ where the pairing gap is exponentially suppressed~\cite{Yang-Gaudin}. This implies that the $q^2$ dependence of the magnetic structure factor should take over only at vanishingly small values of $q$ not reachable in our simulations. In Fig.~\ref{fig5} we compare the low-$q$ behaviour of both structure factors with the linear slope fixed by $S_{\small{D(M)}}(q)=\frac{\hbar q}{2}\sqrt{\frac{\chi_{\small{D(M)}}}{mn_{\text{eq}}}}$ where $\chi_{\small D}=\frac{n_{\text{eq}}}{mc_{\text{eq}}^2}$ is the isothermal compressibility and $\chi_{\small M}=\frac{2}{g+|\tilde{g}|}$ is the estimate of the magnetic susceptibility assuming the spin sector gapless~\cite{note}. 

Coherence properties in the liquid state at equilibrium are characterised by the behaviour of the one-body density matrix (OBDM). This is invariant under the exchange of the two species and is defined as 
\begin{equation}
\rho(s)=\langle\psi_{a(b)}^\dagger(x+s)\psi_{a(b)}(x)\rangle \;,
\label{OBDM}
\end{equation}
in terms of the field operators giving the density of each component: $n_{a(b)}(x)=\psi_{a(b)}^\dagger(x)\psi_{a(b)}(x)$. In systems exhibiting off-diagonal long-range order the OBDM at large distance $s$ reaches a constant value identified with the condensate density. However, Bose-Einstein condensation does not exist in 1D and at $T=0$ the OBDM is expected to decay with a power law. In Fig.~\ref{fig6} we show the results of $\rho(s)$ by varying the ratio $|\tilde{g}|/g$ and we find a clear algebraic decay with the distance: $\rho(s)\propto1/|s|^\alpha$, which sets in at sufficiently large values of $s$. The value of the exponent $\alpha$ is reported in the inset of Fig.~\ref{fig6} as a function of the ratio $|\tilde{g}/g|$. The exponent ranges from very small values at $|\tilde{g}|/g\simeq1$, where the GGP theory is applicable, to values as large as $\alpha\simeq0.3$ close to the critical ratio of coupling constants. We emphasise that in the Tonks-Girardeau regime of a single-component Bose gas, corresponding to particles being impenetrable and behaving like fermions, the OBDM decays as $\rho(s)\propto1/\sqrt{|s|}$~\cite{Lenard64}. The values of $\alpha$ found moving towards the critical ratio $(|\tilde{g}|/g)_{\text{crit}}$ arise from very strong correlations acting between particles of the same species, which result in a suppression of the momentum distribution peak at low wave vectors $n(q)\propto1/|q|^{1-\alpha}$ as entailed by the relation between $n(q)$ and the OBDM via the Fourier transform $n(q)=\int\; ds\; e^{iqs}\rho(s)$.     

In conclusion, we have investigated the properties of the bulk liquid state in attractive 1D Bose-Bose mixtures by using exact QMC methods. We find that the liquid state can exist only if the ratio of coupling constants exceeds a critical value. The thermodynamic properties of the equilibrium state derived from our simulations are crucial ingredients when studying the stability and the collective modes of the liquid droplets realised in experiments. In particular, we find that regimes of strongly correlated liquid states are achievable in 1D well beyond the conditions of applicability of the weak-coupling GGP theory.

{\it Acknowledgements:} This work was supported by the QUIC grant of the Horizon 2020 FET program, by Provincia Autonoma di Trento and by the grant FIS2014-56257-C2-1-P of the MICINN (Spain). G.E.A. thankfully acknowledges the computer resources at MareNostrum and the technical support provided by Barcelona Supercomputing Center (RES-FI-2017-3-0023).

\end{document}